\documentclass[]{emulateapj}
\usepackage[backref, breaklinks, colorlinks, citecolor=blue, linkcolor=magenta]{hyperref}
\usepackage[dvipsnames]{xcolor}

\slugcomment{Submitted to The Astrophysical Journal}

\shorttitle{Gas Accretion in MW-like Galaxies}
\shortauthors{Sanchez et. al.}

\begin{document} 

\title{Preferential Accretion in the SMBH of Milky Way Size Galaxies Due to Direct Feeding by Satellites}

\author{N. Nicole Sanchez \altaffilmark{1, 2, 3}}
\author{Jillian M. Bellovary\altaffilmark{4, 5}}
\author{Kelly Holley-Bockelmann\altaffilmark{2, 3}}
\author{Michael Tremmel\altaffilmark{1}}
\author{Alyson Brooks\altaffilmark{6}}
\author{Fabio Governato\altaffilmark{1}}
\author{Tom Quinn\altaffilmark{1}}
\author{Marta Volonteri\altaffilmark{7}}
\author{James Wadsley\altaffilmark{8}}

\affil{$^1$Astronomy Department, University of Washington, Seattle, WA 98195, US, sanchenn@uw.edu}
\affil{$^2$Department of Natural Sciences and Mathematics, Fisk University, 1000 17th Avenue N., Nashville, TN 37208, USA}
\affil{$^3$Department of Physics and Astronomy, Vanderbilt University, PMB 401807, Nashville, TN 37206, USA}
\affil{$^4$Queensborough Community College, New York, NY 11364, USA}
\affil{$^5$Department of Astrophysics, American Museum of Natural History, Central Park West at 79th Street, NY 10024, USA}
\affil{$^6$Department of Physics and Astronomy, Rutgers, The State University of New Jersey, 136 Frelinghuysen Road,Piscataway, NJ 08854, USA}
\affil{$^7$Institut d’Astrophysique de Paris, Sorbonne Universitès, UPMC Univ Paris 6 et CNRS, UMR 7095, 98 bis bd Arago, 75014 Paris, France}
\affil{$^8$Department of Physics and Astronomy, McMaster University, Hamilton, ON L8S 4M1, Canada}


\begin{abstract}\label{abs:abstractlabel}

Using a new, high-resolution cosmological hydrodynamic simulation of a Milky Way-type (MW-type) galaxy, we explore how a merger-rich assembly history affects the mass budget of the central supermassive black hole (SMBH). We examine a MW-mass halo at the present epoch whose evolution is characterized by several major mergers to isolate the importance of merger history on black hole accretion. This study is an extension of Bellovary et. al. 2013, which analyzed the accretion of high mass, high redshift galaxies and their central black holes, and found that the gas content of the central black hole reflects what is accreted by the host galaxy halo. In this study, we find that a merger-rich galaxy will have a central SMBH preferentially fed by gas accreted through mergers. Moreover, we find that the gas composition of the inner $\sim$ 10 kpc of the galaxy can account for the increase of merger-accreted gas fueling the SMBH. Through an investigation of the angular momentum of the gas entering the host and its SMBH, we determine that gas accreted through mergers enters the galaxy halo with lower angular momentum compared to smooth accretion, partially accounting for the preferential fueling witnessed in the SMBH. In addition, the presence of mergers, particularly major mergers, also helps funnel low angular momentum gas more readily to the center of the galaxy. Our results imply that galaxy mergers play an important role in feeding the SMBH in MW-type galaxies with merger-rich histories.

\end{abstract}



\section{Introduction}\label{sec-intro}
Supermassive black holes (SMBHs) are thought to exist in almost all massive galaxies \citep[see][for a review]{Kormendy2013}. In the canonical picture of BH growth, these black holes may become active galactic nuclei (AGN) during periods of high accretion and wane in periods of quiescence \citep{Begelman1980,Alexander2005,Papovich2006,Volonteri2012}. The host galaxy's size, star formation rate, and other environmental effects may help to influence the growth of the black hole residing at its center; however, there are still uncertainties concerning the relationship between these SMBHs and their much larger host galaxies, as well as how they grow and evolve together \citep{Haehnelt2000,DiMatteo2005,Hopkins2006,Fu2008,Sijacki2009,Silverman2009,Micic2011,Mullaney2012}.

The M$-\sigma$ relation, which relates the SMBH's mass and the velocity dispersion of the host galaxy's central stellar population, gives some insight into the complex interplay between these objects \citep{Ferrarese2000,Gebhardt2000}. A prominent trend appears, as SMBHs tend to scale with the velocity dispersion of the host galaxy bulge. The tightness of the relation is significant and can be seen over several orders of magnitudes in velocity dispersion and black hole mass \citep[e.g.][]{Merritt2001,Tremaine2002,Graham2011,Mcconnell2013}. Scatter exists among the low mass galaxies and a deviation may appear at the high mass end, where overmassive BHs may reside \citep{VanDenBosch2007,Moster2010,Natarajan2011,Emsellem2011,Volonteri2016a}. However, scatter in less massive galaxies may imply that there are several channels of black hole growth at play in the low mass end of the relation \citep{Micic2007,Volonteri2009,Reines2013,Graham2014}. One standard explanation for the M$-\sigma$ relation lies in galaxy mergers, which build up galaxies, feed SMBHs, and assemble bulges \citep[e.g.][]{DiMatteo2005,Shen2008}. Major mergers are thought to supply gas to the central SMBH resulting in feedback which quenches star formation and affects the structure of the galaxy \citep{Schawinski2010}.

Major mergers between massive galaxies are thought to be efficient fueling mechanisms for bright AGN. Additionally, the most massive, highest-luminosity AGN (i.e. quasars) reside in incredibly luminous infrared galaxies where star formation is abundant, signifying that major mergers may have recently occurred \citep{Treister2012}. Distorted morphologies are often characteristics of quasar hosts, and companions can also be present around quasars, both of which are evidence that strengthen the possibility of a recent merger having affected their lifetimes \citep{Ellison2010}.

These major mergers can also strongly disturb gas-rich galaxies producing resonant tidal torques that allow large influxes of material to funnel directly into the center, fueling bursts of star formation and SMBH accretion \citep[e.g.][]{Sanders1988,Sanders1996,Barnes1991,Mihos1996,Hopkins2006,Richards2006,Reddy2008,Hopkins2010}. At small scales closer to the SMBH, the larger tidal torques from a major merger less effectively drive gas into the inner most region of the galaxy; however, perturbations at all scales from the merger could still drive accretion into the smaller inner region of galaxy, though the rapid decay of these perturbations may not encourage gas flow. Other large scale instabilities such as bars and spiral waves are also proficient fueling mechanisms for funneling gas into the galaxy; however, these cases can inhibit small scale gas accretion through other complications. While there is still some uncertainty regarding the processes that transport gas through last $\sim$ 1 kpc to the SMBH, \cite{Hopkins2010} has shown that major mergers between gas-rich galaxies can result in non-axisymmetric gravitational instabilities which can drive BH accretion within the inner most $\sim$ 0.1 pc.

In many less massive and less luminous AGN, however, there is a clear lack of distorted morphology, close neighbors, and/or other obvious merger evidence \citep{Ryan2007,Schawinski2011,Ellison2013,Hicks2013}. It is also important to note that many of these AGN exist in spiral galaxies, which are unlikely to have been recently disturbed by major mergers \citep{Schawinski2011,Kocevski2011}. Nevertheless, some evidence suggests that disturbed galaxies may reform a disk quickly, even after a major merger, as long as it is gas-rich \citep{vanGorkom1997}. The rapid disk reformation of the galaxy in this paper was previously studied by \cite{Governato2009} (See \ref{results}). More recently, \cite{Treister2012} has suggested that only the highest luminosity AGN require fueling via major mergers; $\sim$ 90 $\%$ of AGN across all redshifts are fueled by various other mechanisms, which may include minor mergers, flybys, and smooth accretion, whereby gas is directly accreted via large filaments from the ambient intergalactic medium \citep{Cox2006,Bellovary2013,Sinha2012,Dubois2012,DiMatteo2016}.

Smooth accretion, in particular, may play an important role in fueling these low mass galaxies. Halos less than 10$^{11}$ $M_{\odot}$ can accrete filaments of unshocked gas; thereafter, gas will shock heat to the virial temperature of the halo \citep{Keres2005}. Even for massive halos, unshocked gas may still penetrate shocked regions to fuel the galaxy \citep{Brooks2009,Dekel2009,Nelson2013}. In addition, SMBH feedback, the depositing of energy and momentum back into the gas reservoir during accretion, also affects the overarching structure of the host galaxy \citep{Governato2009a}. Secular processes, including bar formation and disk instabilities, may also be prominent forms of accretion for these SMBHs \citep{Athanassoula2016,Kormendy2013}. 

It is clear that galaxy hosts grow through a variety of channels that depend on mass, environment, and interaction history. Therefore, we want to understand how these different galaxy evolutionary paths translate into SMBH fueling mechanisms, and see how they affect the fueling gas flowing into the SMBH itself. \cite{Bellovary2013} (hereafter, B13) compared simulations of three high mass, high redshift galaxies and found that while mergers and smooth accretion both efficiently build up galaxies, no particular dynamical process was more adept at feeding the SMBH. However, with only minor mergers, these galaxies represented relatively quiet merger histories. Using a similar method as B13, this work examines the SMBH and galaxy fueling mechanisms of a MW-mass galaxy with a rich merger history. MW-type galaxies host SMBHs on the order of 10$^6$ $M_{\odot}$, which are likely the most common type of massive black hole, yet little is known about them or how they may grow \citep{Kormendy2013}. Through this examination, we hope to better understand the coevolution of SMBHs and their hosts in this class of galaxy. 

In this study we analyze the Milky Way-type galaxy, h258, which has a history characterized by major mergers. Since this galaxy is similar to the MW in virial mass, stellar mass, and circular velocity, without a deeper examination, we may not recognize the turbulent history from which it results. We will pinpoint the origins of gas entering the SMBH and halo to look for clues about SMBH fueling within this galaxy. By examining its assembly, and its SMBH's fueling, we can determine the accretion rate and gain further understanding about how SMBHs grow over a range of merger histories in galaxies like our own. 



\section{Simulation Parameters}\label{sec-model}

Using the smoothed particle hydrodynamics (SPH) N-body tree code, Charm N-body GrAvity solver \citep[ChaNGa;][]{Menon2015}, we ran an initial dark matter-only, uniform resolution volume of 50 h$^{-1}$ Mpc on a side to identify a MW-mass halo at z = 0 for further examination. This DM-only simulation assumed WMAP 3 parameters \citep{Spergel2007}: $\Omega_m$ = 0.24, $\Omega_{\Lambda}$ = 0.76, $H_0$ = 73 km/s, and $\sigma _8$ = 0.77. Halo h258 was chosen for its Milky Way-mass at z=0 and its active merger history. The halo has a virial mass of $M_{\rm vir} = 8.6 \times 10^{11} M_{\odot}$ at z = 0 defined relative to a critical density, $\rho_c$, where $\rho / \rho_c$ = 200, and the virial radius is defined as the radius which encloses a density 200 times that of $\rho_c$. Two recent major mergers characterize the h258 halo at z=1.8 and z=1.2. We constructed a ``zoom-in'' high resolution simulation on this galaxy, including gas and star particles, using the volume-renormalization of \cite{Katz1993}, resimulating only a few virial radii from the main halo at the highest resolution from z=150 to z=0.  

We note that a lower resolution version of h258 was run using the Gasoline code \citep{Wadsley2004}. Our higher resolution h258 run has a spline force softening length of 174 pc and initial gas particle masses of 2.7 $\times$ $10^4 M_{\odot}$. Star particles are created with 30$\%$ of their parent gas particle mass, allowing a mass of 8100 $M_{\odot}$. Halo h258 contains about 5 million DM particles inside the virial radius at z=0 and over 14 million DM, star, and gas particles total. The resolution of both force and mass in these simulations is comparable to the ``Eris'' simulation which has one of the highest resolutions for an N-body+SPH cosmological simulation of a Milky Way-mass galaxy so far produced \citep{Guedes2011}.  

Compared to the previous h258 simulation, the ChaNGa simulated h258 scales better and includes a new improved SPH formalism \citep{Keller2014}. The hydrodynamic treatment now includes a geometric density average\textemdash (P$_i$ + P$_j$)/($\rho_i \rho_j$) rather than P$_i$/$\rho_{i}^2$ + P$_j$/$\rho_{j}^2$ where P$_i$ and $\rho_i$ are the particle's pressure and density\textemdash in the force expression, in addition to the standard SPH density estimator \citep{Ritchie2001}. Adjusting the force expression diminishes the numerical surface tensions due to shear flows, such as Kelvin-Helmholtz instabilities. We also apply a consistent and entropy-conserving energy equation to account for the modified force expression and correctly model strong shocks, such as Sedov blasts.

Our simulation introduces a uniform UV background at z$\sim$9 to simulate the cosmic reionization energy using the formula of \cite{Haardt2012}. To model star formation, gas particles can stochastically spawn up to 3 stars with a star formation efficiency parameter of c$^*$ = 0.1 once the density threshold and temperature satisfy conditions for star formation (10.0 amu cm$^{-3}$; T \textless 10$^4$ K). As shown in \citet{Governato2009a}, this high density threshold is necessary to produce bursty star formation events in the high-density peaks of the ISM. Because we are able to resolve gas smoothing lengths up to 10 times smaller than the gravitational softening length, we are confident that these high density peaks can be properly tracked and resolved. If all the criteria are met, the probability a gas particle will form a star is given by

\begin{equation}
p = \frac{m_{\rm gas}}{m_{\rm star}} (1 - e^{-c^*\Delta t/t_{\rm form}})
\end{equation}

where $m_{\rm star}$ and $m_{\rm gas}$ are the star and gas particle masses, $t_{\rm form}$ is the gas particle's dynamical time, and we set the time between star formation episodes, $\Delta t$, to 1 Myr. \cite{Bellovary2011} describe this star formation criteria; however, they mistakenly exclude the negative sign in the exponent, and we present the corrected version here.

Realistic star formation histories result in galaxies which have realistic density profiles and lie on the observed scaling relations (mentioned in detail below).  The lower-resolution simulations mentioned in this paper have a lower threshold, since the high density peaks are not resolved. These star particles represent a Kroupa initial mass function \citep{Kroupa1993}.  Molecular hydrogen and metal-line cooling are not included, though we implement a low-temperature extension to the cooling curve to trace metals \citep{Bromm2001} and the metal diffusion prescription of \cite{Shen2010}. We do not expect the omission of metal cooling to affect our results. Gas which is smoothly accreted onto a galaxy is expected to have low metallicity, so cooling by metals (for example, after gas undergoes a shock) will have low efficiency in this instance. According to \cite{Christensen2014}, broad galaxy properties such as rotation curves, star formation histories, and density profiles are consistent across various cooling models; additionally, the primordial cooling model we use here is in excellent agreement with many properties of the more sophisticated molecular ISM model. The effect of metal cooling on the categorization and subsequent evolution of the gas would be negligible.

Supernova (SN) feedback releases $10^{51}$ ergs of thermal energy within a ``blastwave'' radius determined by the equations of \cite{Ostriker1988}. In the affected region, cooling turns off for a time relative to the expansion phase of the SN remnant also determined by the blastwave equation. SN Ia and II rates from \cite{Thielemann1986} and \cite{Woosley1995}, respectively, are implemented through the \cite{Raiteri1996} method, which uses the stellar lifetime calculations of the Padova group \citep{Alongi1993, Bressan1993, Bertelli1994} to describe stars with varying metallicities. Both the supernova ``blastwave'' radius and supernova (Ia and II) prescriptions are described in detail by \cite{Stinson2006}. While it's true that a different treatment of the ISM and SNe feedback might alter the structure of the ISM, \cite{Christensen2014} examined simulated spiral and dwarf galaxies utilizing similar SNe prescriptions and three different ISM models, and determined that the resulting ISM remained consistent with each other. Additionally, our galaxy is in good agreement with galaxies affected by superbubble SN feedback \citep{Keller2014}.

Simulated galaxies are shown to conform with the observed Tully-Fisher relation \citep{Governato2009}, the size-luminosity relation \citep{Brooks2011}, and the mass-metallicity relation \citep{Brooks2007,Christensen2015}, in addition to having realistic matter distributions and baryon fractions \citep{Governato2009a,Guedes2011}. The parameter and resolution choices described above allow the galaxies to adhere to the stellar-mass-halo-mass relation at z=0 and maintain a realistic period of star formation \citep{Moster2010,Munshi2013,Brooks2007,Maiolino2008}. Given that the simulations are in accordance with observations, we are confident that it reasonably represents growth in the galaxy and its SMBH.
 
Since there are uncertainties in black hole seed formation, we model BH seeding that is broadly consistent with several theories of direct collapse black holes \citep{Couchman1986, Abel2002, Bromm2004} and Population III stellar remnants \citep{Loeb1994, Eisenstein1995, Koushiappas2004, Begelman2006, Lodato2006}. While this method allows the BH formation process to remain physically motivated, BH seeds form if their parent gas particle matches the criteria required for star formation and also maintains zero metallicity, a requirement of many direct collapse models. A probability of $\chi_{\rm seed}$ $\sim$ 0.1 is applied to determine whether a gas particle (with the above specifications) will become a BH seed with the same mass as its parent gas particle. This probability was chosen to match the predicted occupation fraction of BH seeds at z $\sim$ 3 \citep{Volonteri2008}. However, BH particles cease to form once the global metallicity increases due to star formation, resulting in BH seeds only forming early on in the simulation (See \cite{Bellovary2011} for further details).

In many occasions, seed BH particles form one at a time. Feedback from accretion begins immediately after formation, preventing further massive black holes (MBHs) from forming nearby due to the increased temperature. On some occasions, however, more than one MBH seed can form in the same location at the exact same time, because multiple particles meet the formation criteria.  In this instance, they often merge quickly.  Since we do not resolve the direct collapse process, we do not consider this multiple-seed formation physical; rather, one can consider it a crude form of an initial mass function.  The direct collapse seed mass is not well constrained, and the resulting seed acts the same as a single-seed predecessor does throughout the remainder of the simulation.

The requirement that BH seeds must form from zero metallicity gas particles also causes BH formation to be confined in areas of primordial star formation in the earliest and most massive halos in the simulation. In this technique, BH formation is dependent only on local environment, neglecting any large-scale properties of the host halo. We use the sub-grid prescription for modeling the effects of dynamical friction on SMBH orbits from \cite{Tremmel2015}, which has been shown to produce realistic sinking times for SMBHs.  This prescription, combined with our high resolution to minimize two-body interactions and numerical noise, results in SMBHs that can remain stable at galactic center while also, when appropriate, experiencing realistic perturbations and sinking timescales during and after galaxy interactions and mergers \citep{Bellovary2011}.

Black hole mergers occur when they are separated by less than twice the softening length and satisfy $(1/2) \delta v^2 < \delta a \cdot \delta r$ (which is an approximation of being gravitationally bound),  where $\delta v$ and $\delta a$ are the velocity and acceleration differences between the two black holes and $\delta r$ is the distance separating them. In addition to gaining mass via mergers, black holes accrete through the Bondi-Hoyle mechanism:
\begin{equation}
\dot{M} = \frac{4 \pi \alpha G^2 M^{2}_{\rm BH} \rho}{(c^{2}_{s} + v^2)^{3/2}},
\end{equation}
where $\alpha$ is a constant of order 1, $\rho$ is the density of the surrounding gas, $c_s$ is the sound speed, $v$ is the black hole's relative velocity to the gas, and the accretion rate is Eddington-limited. Feedback is applied to surrounding gas with an energy boost determined by the accreted mass as follows: $\dot{E}$ = $\epsilon _{r}$$\epsilon_{f}$$\dot{M}$$c^2$ where $\dot{M}$ is the accreted mass, and $\epsilon _r = 0.1$ and $\epsilon _f = 0.03$ are assumed for the radiative efficiency and feedback efficiency, respectively. This energy is distributed as thermal energy to the 32 nearest particles via a kernel probability function. Though other groups use a higher value for feedback, $\epsilon _f = 0.05$ \citep{Sijacki2007,DiMatteo2008}, we find that $\epsilon_f = 0.03$ in our code produces MBHs in better agreement with MBH-host galaxy scaling relations. However, as our main concern is in the relative proportion of gas from various origins (See \ref{results}) and we restrict our analysis of the angular momentum of gas to only the timestep of entry into the main halo, our results are not sensitive to our choices of $\epsilon _{r}$ or $\epsilon_{f}$. This same model was additionally used by B13 at a lower resolution and without the addition of the dynamical friction prescription. While the black hole seed and accretion models may determine the final mass of the SMBH, by comparing the relative proportions of gas accreted by the black holes, we avoid these model dependencies. Additionally, because gas is categorized as it enters the outskirts of the galaxy, far from any SMBH, the relative fractions accreted by the central black holes are not affected by the accretion and feedback models.



\section{Simulation Analysis}\label{redux}

We first identify halos using the Amiga Halo Finder which uses an overdensity criterion for a flat universe \citep{Knebe2001,Knollmann2009,Gill2004} to set the virial radius in the primary halo. We select the primary halo to be the most massive at z=0 in the high resolution region. The central SMBH in the primary halo has a mass of $1.3 \times 10^{7} M_{\odot}$ and a velocity dispersion in the bulge of $\sigma$ $\sim$ 152 km s$^{-1}$, indicating that h258 lies on the M-$\sigma$ relation. Additionally, the disk scale length is ~2.8 kpc, comparable to that of the MW, while the total halo mass, M$_{DM}$ = 8.6 $\times$ 10$^{11} M_{\odot}$, and stellar mass, M$_*$ = 5.5 $\times$ 10$^{10} M_{\odot}$, show that the galaxy halo fits on the SMHM relation using the correction factors from \cite{Munshi2013}.

In this analysis, we retrace each gas particle that was accreted by the halo or SMBH, following the gas back through its journey in the halo and recording its host halo and time of accretion \citep{Brooks2009}. The particles are then classified into types by their method of entrance into the primary halo. Gas that belonged to a different halo than the primary prior to accretion is classified as ``clumpy,'' entering the primary halo through mergers. All other gas is classified as ``smooth'' accretion, and is then subdivided into two categories: ``unshocked'' and ``shocked.'' Unshocked gas will usually flow into the halo via large-scale, dark matter filaments \citep{Keres2005,Bellovary2013}. It is possible for unshocked gas to be dense enough to pierce an already developed shock, allowing it to funnel into the galaxy core where it can be accreted onto the SMBH \citep{Nelson2013}. 

However, as we discussed in Section 1, if the galaxy halo is $\gtrsim$ 10$^{11}$ M$_{\odot}$, the gas is known to shock-heat to the virial temperature of the halo. We identify shocked particles through an increase in entropy and temperature using the following criteria:
\begin{equation}
T_{\rm shock} \geq 3/8 T_{\rm vir},
\end{equation}
where T$_{\rm vir}$ is the virial temperature of the halo, T$_{\rm shock}$ is the temperature of the gas particle, and the minimum change in entropy is
\begin{equation}
\Delta S \geq S_{\rm shock} - S_0,
\end{equation}
where S$_0$ is the initial entropy of the gas particle, and 
\begin{equation}
S_{\rm shock} = log_{10}[(3/8 T_{\rm vir})^{1.5}/4 \rho_0],
\end{equation}
where $\rho_0$ is the gas density prior to encountering the shock. Therefore, gas particles must reach both an entropy and temperature threshold to be labeled as ``shocked,'' and all smoothly accreted gas which does not is labeled as ``unshocked''. Additionally for a ``shocked'' classification, the gas must be entering the virial radius and a minimum galaxy halo mass ($\gtrsim$ 10$^{11}$ M$_{\odot}$) must be reached \cite[see][for further details]{Brooks2009}. Since our halo is $\sim$ 10$^{12}$ M$_{\odot} $ by z = 0, we should expect to find more shocked gas entering the halo at later times. Both types of smoothly accreted gas are tracked from the moment they enter the virial radius until they reach a cutoff radius at 10$\%$ of the virial radius (0.1 R$_{vir}$) at which point supernova feedback may appear as virial shocking and cause contamination in our estimates of shocked accretion. Thus, gas may be labeled as shocked if it meets our criteria between the times when it crosses R$_{vir}$ and when it reaches 0.1 R$_{vir}$. This cutoff also accounts for any AGN feedback we might encounter.

Once all the gas particles have been individually categorized, we can use these labels to classify the gas accreted by the SMBH, and we can better contrast the processes that feed the galaxy halo and its SMBH in MW-mass halos.


\begin{figure}
\centerline{\resizebox{0.75\hsize}{!}{\includegraphics[angle=0]{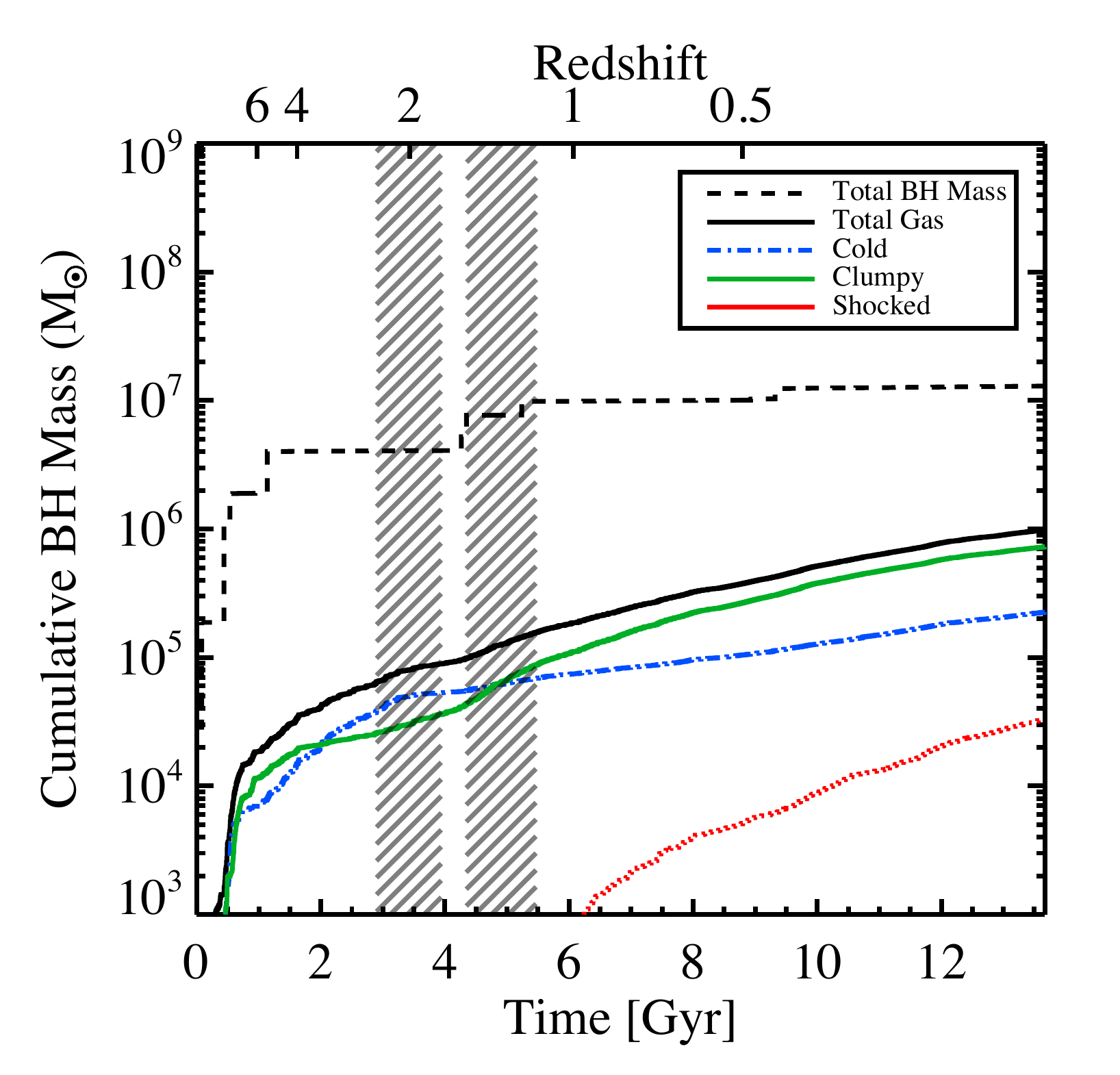}}}
\caption[]{The central BH's cumulative mass as a function of time and redshift. The black dashed line indicates the total cumulative BH mass. The black solid line indicates the total accreted gas mass. The blue dot-dashed line indicates smoothly accreted gas mass that remains unshocked after entering the virial radius of the main halo. The green solid line indicates the gas mass accreted through mergers. The red dashed line indicates accreted gas mass that was shocked upon entry into the halo. Gas tracking begins when the BH exists in a galaxy halo of a non-diminutive size, though the BH did exist and merged with other BHs ($\sim$ 5) prior to this point. Major mergers are marked with grey hatched regions.}
\label{hrh258allmassgas} 
\end{figure}

\begin{figure}
\centerline{\resizebox{0.75\hsize}{!}{\includegraphics[angle=0]{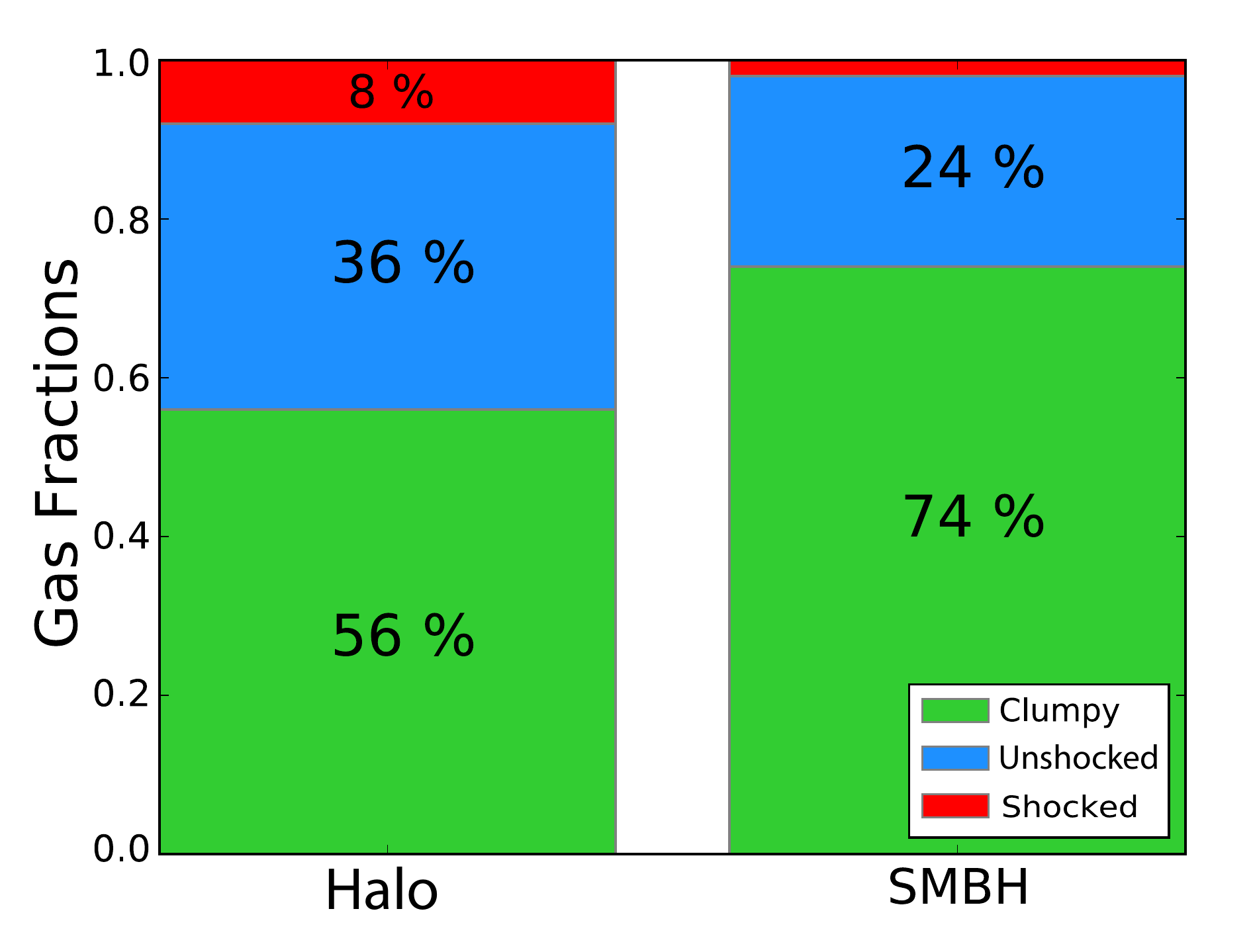}}}
\caption[]{Gas fractions of the gas particles accreted by the main halo (left) and the SMBH (right), distinguished by type. Blue, green, and red distinguish gas gained via smooth accretion that remains unshocked, gas gained through mergers, and smoothly accreted gas that is shocked upon entry, respectively.}
\label{hrh258stackfrac}
\end{figure}


\section{Results} \label{results}


The galaxy h258 is characterized by two major mergers; the first occurs at z $\sim$ 1.8 (mass ratio, q $\sim$ 0.8) and the second at z $\sim$ 1.2 (q $\sim$ 1). Despite its merger-rich history, gas accretion smoothly increases the cumulative black hole mass in h258 throughout its evolution as can be seen in Figure \ref{hrh258allmassgas}. The black dashed line in Figure \ref{hrh258allmassgas} indicates the total cumulative BH mass (including both mass from gas and BH mergers), while the black solid line indicates the total accreted gas mass. The blue dot-dashed line represents the gas mass accreted via unshocked gas, while the green solid line and red dashed line show the gas mass accreted through mergers and shocked gas, respectively. Major mergers are indicated by grey hatched regions. 

It is important to point out that the largest part of the mass budget is not gas at all, but other black holes that have merged with the SMBH seed; the final distribution of mass in the SMBH ($1.3 \times 10^{7} M_{\odot}$) comes primarily ($\sim$90 $\%$) from mergers with other black holes. This has important implications for gravitational wave astronomy, increasing the event rate for SMBH assembly at high redshifts \citep{Holley-Bockelmann2010}.

Although the intent of this paper is to focus on the origin of the gas accreted by the SMBH, it is worthwhile to examine the remainder of the SMBH’s growth, which consists of black hole mergers.  While the contribution from other seed SMBHs appears significant, we point out that the uncertainties in seed masses and formation efficiencies results in a large uncertainty in the exact mass that may be acquired by SMBH mergers.  Additionally, there are repercussions regarding gravitational waves that we do not consider here as well, such as recoil upon merging. We leave a treatment of the dark side of the SMBH mass budget to a future paper, as it calls for a statistical or semi-analytic approach to incorporate the effects of seed model, black hole spin, and gravitational wave recoil.

Aside from this significant BH assembly, the largest gain in accreted SMBH gas mass comes from gas accreted through mergers after z $\sim$ 1. From Figure \ref{hrh258allmassgas}, we see that gas accreted through mergers makes up the majority of accreted mass entering the SMBH at early times; however, the transition between when smooth, unshocked accretion and gas accreted from mergers dominates are clearly distinguished. While clumpy gas (green) dominates gas mass accretion in the SMBH at the earliest time, unshocked gas (blue) overtakes it for a short time before clumpy gas once again dominates by z $\sim$ 1.5.

This low redshift transition to a clumpy gas preference results in the large fraction of clumpy gas seen in the accreted mass fractions in the SMBH (Figure \ref{hrh258stackfrac}). Figure \ref{hrh258stackfrac} depicts the fractions of total gas accretion in the galaxy halo and the SMBH at z=0, again differentiated by gas origin. The gas accreted by the halo is half (56 $\%$) comprised of gas accreted through mergers, with 36 $\%$ of the gas entering through unshocked, smooth accretion. The smallest fraction of the total gas is comprised of shocked gas (8 $\%$ ). Unlike the halo, nearly three quarters (74 $\%$) of the gas accreted by the central SMBH was accreted via mergers, while only a quarter (24 $\%$) is comprised of unshocked, smoothly accreted gas. Shocked gas makes up the last 2$\%$ of total gas entering the SMBH. \textit{It is evident then that the SMBH more readily accretes gas gained through mergers.} While this result is consistent with the work of \cite{Dubois2015} which explores how galaxy mergers may be necessary for triggering black hole growth in low mass galaxies, it is contrary to B13 which found that in high redshift, high mass galaxies, the fractions of gas comprising the SMBH and its host were nearly the same. While the study by B13 focuses on high mass galaxies at high redshift, they all have similar gas fractions and masses (total masses on the order of $10^{11} M_{\odot}$ and gas masses of a few $10^{10} M_{\odot}$) compared to the immediate progenitors of h258 (which merge at z $\sim$ 1.2 and z $\sim$ 1.8). In addition, the smooth accretion histories are broadly similar, in that each host galaxy forms a shock front and begins accreting shocked-mode gas about half way through its evolution ($\sim$ 6 Gyr for h258, compared to $\sim$ 1 Gyr for the B13 galaxies).  The similar smooth accretion histories, masses, and gas fractions do point to the gas of merger origin being the key factor in the difference between our results and those of B13. We suggest that when major mergers are a key part of a galaxy's assembly history, these mergers may also drive SMBH growth.

\begin{figure}
\centerline{\resizebox{0.85\hsize}{!}{\includegraphics[angle=0]{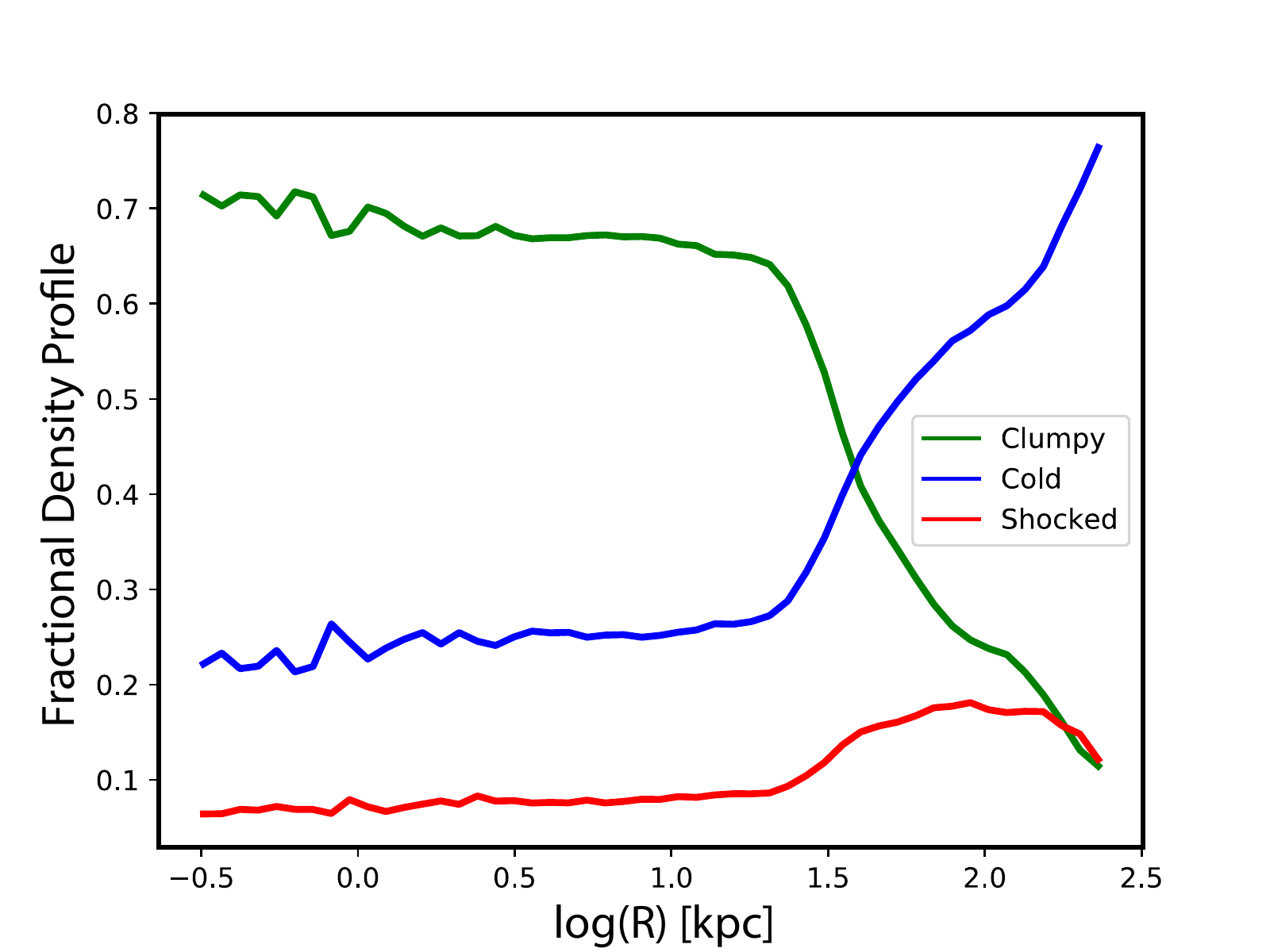}}}
\caption[]{Fractional radial density profiles of the gas that enters the main halo by z=0. All our lines add up to 1 in each radial bin. Gas within the galaxy halo's central 10 kpc is comprised of gas with similar fractions of clumpy, unshocked, and shocked gas compared to that which is accreted by the SMBH (Figure \ref{hrh258stackfrac}). The green, blue, and red lines indicate clumpy, unshocked, and shocked gas, respectively.}
\label{densityprofile} 
\end{figure}

A previous study by \cite{Fu2007} supplies direct observational evidence that gas from a merger can be funneled directly into the vicinity of the SMBH (\textless 1 pc). Their study examined a collection of twelve low-redshift quasars, half of which are characterized by luminous extended emission-line regions (EELRs). These EELRs were found to have metallicities below the mass-metallicity correlation of normal galaxies and are thought to have resulted from massive, galactic superwinds accompanying the creation of the powerful radio jets associated with the quasars. The quasars hosting EELRs were also found to have low-metallicity broad-line regions (BLRs) at their centers, while the quasars without EELRs hosted BLRs with metallicities above Z$_{\odot}$. \cite{Fu2007} determined that the presence of these low-metallicity EELRs as well as the similar, low-metallicity BLRs are evidence of a merger with a gas-rich galaxy. Such an interaction may explain both the infusion of low-metallicity gas into the BLRs and the subsequent ejection of that gas by the radio jets, forming the EELRs. \cite{Fu2008} do not directly state that the AGN activity was powered by gas directly from the merging galaxy; however, they do infer that the low-metallicity BLR and EELR regions originate directly from the interloper.  Therefore it is quite likely that at least some of the gas from the merging galaxy, which seems to venture quite near the SMBH, may be accreted onto it in non-negligible amounts. We specifically traced the fraction of gas accreted via mergers that enters the galaxy (\textless 10 kpc from the center) and determined this clumpy gas fraction is comparable to that accreted by SMBH. Figure \ref{densityprofile} describes the fractional radial density profiles of the gas in the halo where clumpy, unshocked, and shocked gas are shown in green, blue, and red, respectively. The composition of the gas nearest the SMBH can account for increase we see in the amount of clumpy gas comprising the gas mass of the SMBH over that of the halo. While our study did not measure metallicity or include radio jets, our overall results support the conclusion that gas from incoming galaxy mergers with initially low angular momentum can efficiently lose further angular momentum through dynamical interactions with the larger galaxy, and can therefore be more readily accreted by the SMBH.

\begin{figure}
\centerline{\resizebox{0.75\hsize}{!}{\includegraphics[angle=0]{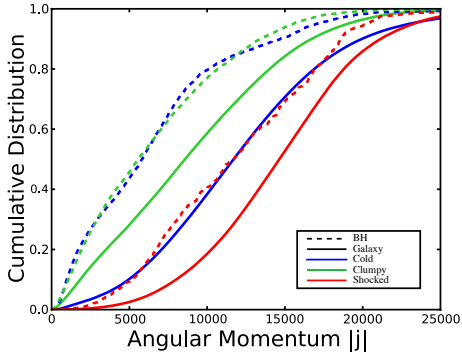}}}
\caption[]{ Normalized cumulative distribution of angular momentum (kpc km s$^{-1}$) of the gas particles accreted onto h258 by z = 0.  Gas particles accreted onto the main halo (solid lines) and central black hole (dashed lines). The green, blue, and red lines indicate clumpy, unshocked, and shocked gas, respectively.}
\label{hrh258angmom} 
\end{figure}

\begin{figure}
\centerline{\resizebox{0.75\hsize}{!}{\includegraphics[angle=0]{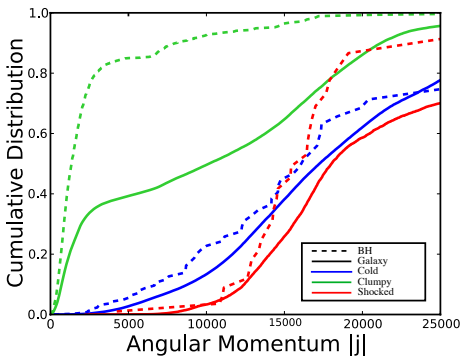}}}
\caption[]{ Normalized cumulative distribution of angular momentum (kpc km s$^{-1}$) of the gas particles accreted onto the h258 halo at a single timestep (z $\sim$ 1.2) during the major merger's greatest influx of accreted gas. About 450,000 gas particles accreted onto the halo at this timestep. Angular momentum of gas particles accreted onto the main halo and central black hole are distinguished by solid and dashed lines, respectively. The green, blue, and red lines indicate clumpy, unshocked, and shocked gas, respectively.}
\label{hrh258angmom_merger2} 
\end{figure}

To better understand the apparent preference for merger-accreted gas, we examine the angular momentum of the gas at the moment it enters the halo. Figure \ref{hrh258angmom} shows a cumulative distribution of the angular momentum of the gas as it enters the halo (solid lines). We further distinguish the gas that enters the SMBH (dashed lines), still considering its angular momentum at the moment of halo entry. The gas is again distinguished by origin (clumpy, unshocked, shocked being green, blue, and red, respectively). We find that the angular momentum of gas entering which eventually reaches the SMBH is lower overall, and that the lowest angular momentum gas is comprised of both clumpy and unshocked gas. This result can be seen in Figure \ref{hrh258angmom_merger2}, which shows the cumulative distribution of the angular momentum of the incoming gas particles at the time of the greatest influx of accreted gas during the merger at z $\sim$ 1.2. (Colors and linestyles as in Figure \ref{hrh258angmom}.) Figure \ref{hrh258angmom_merger2} explicitly shows that the gas ending up in the SMBH enters with the lowest angular momentum. Thus a fraction of the gas accreted via the mergers that characterize this galaxy's evolutionary history may have low enough angular momentum to be efficiently channeled into the center of the primary galaxy.  This gas must lose further angular momentum to be efficiently accreted by the SMBH, which likely occurs via the standard picture of gas in the host galaxy losing angular momentum due to torques from the merger dynamics \citep{Capelo2015}. It is important for us to highlight that since the presumed SMBH accretion method takes place at smaller scales than our resolution limit, we are unable to track the angular momentum loss of the gas through the last few parsecs.

A lower resolution version of the galaxy h258 was run using the N-body code, Gasoline, like the simulations in B13. Despite fundamental differences in the hydrodynamic implementation and gas physics included and the absence of a dynamical friction prescription, an analysis of this low resolution h258 results in a SMBH with the same distinct preference for accreting clumpy gas. A second galaxy, h277, was also run using Gasoline; however, this galaxy was characterized by a quiescent merger history (no major mergers) and resulted in a SMBH whose accreted mass fractions mirrored the halo (as seen in B13). The broad consistency between the low and high resolution simulation of the same galaxy, and the similar results of a quiescent galaxy to the previous study, indicates that the large scale gravitational dynamics could be a main driver of the SMBH fueling in this case. We also stress that that while major mergers may not be the only physical mechanisms by which gas can be funneled into the centers of galaxies (the previous study being a strong example of this), mergers between galaxies clearly play an important role when considering the gas accretion of SMBHs.

\section{Conclusion}
This study examines the gas accretion onto the fully cosmological simulation of a Milky Way-size galaxy to redshift z $=$ 0, with major mergers characterizing its past. We trace the gas into the SMBH at its center and differentiate the gas accreted onto the galaxy halo and SMBH by origin. Gas gained through mergers is classified as ``clumpy'' gas and smoothly accreted gas is separated into ``shocked'' and ``unshocked'' categories. Our goal is to determine what types of gas are primarily feeding the SMBH and the halos of this galaxy class, and to determine what effects the merger history of a galaxy may have on these processes.

A previous study by \cite{Bellovary2013} analyzed high mass, high redshift galaxies and found the gas composition of the SMBHs mirror their host halos. Contrary to these previous results, when we examined a galaxy with an active merger history, we determined that the SMBH at the center more readily accretes gas gained through mergers. This remained true both in an older low resolution simulation of the same galaxy as well as this current iteration. \textit{In both the low and high resolution cases, we see a significant increase in the clumpy gas accreted by the SMBH compared to its host.} We also note that in the high resolution case, we can attribute the increase of clumpy gas to the mergers that characterize h258 and lead to a concentration of this gas residing within the galaxy's inner $\sim$ 10 kpc.

The angular momentum of the accreted gas as it enters the galaxy halo sheds some light on the mechanism driving this preferentially accreted clumpy gas. Smoothly accreted gas, which enters the halo with a wide range of angular momentum, is likely to adjust to the net angular momentum of the halo gas; however, some studies have found that low angular momentum gas can fuel the central SMBH via filaments of smoothly accreted gas \citep{Dubois2012,DiMatteo2016}. Meanwhile, gas entering through mergers can fall to the galaxy's center with minimal interaction with the halo gas. This restriction gives merger-accreted gas the advantage of falling more readily to the center and accreting onto the SMBH. Considering all origins of gas, our study is the first to see a clear contribution to gas from merging galaxies directly falling to the center of the primary galaxy and feeding the SMBH.

While the examination of this single, extreme case of a galaxy with an active merger history depicts a class of galaxy with varying SMBH accretion methods, a further study of cases with varying merger histories is required to begin understanding the broad spectrum of Milky Way-mass galaxy accretion \citep{Pontzen2016}. Additionally, examinations of other extreme case, e.g. galaxies with varied but still merger-rich histories, may strengthen the validity of this result. Through this study, we show that the presence of major mergers can play an important role in the final compositions of central SMBHs, but the question of how important these mergers are remains to be seen.
 

\acknowledgments
Resources supporting this work were provided by the NASA High-End Computing (HEC) Program through the NASA Advanced Supercomputing (NAS) Division at Ames Research Center. Results were partially obtained using the analysis software Pynbody (https://github.com/pynbody/pynbody). We thank the Fisk-Vanderbilt Masters-to-PhD Bridge program for the funding and support of this research. We also wish to thank the anonymous referee for their thorough reading of the manucript and their useful comments which improved its quality. J.B. acknowledges generous support from the Helen Gurley Brown Trust. M.V. acknowledges support from NASA award ATP NNX10AC84G. 

\bibliography{./Sanchez2016.bib}

\end{document}